\documentclass[10pt,amsmath]{iopart}
  \expandafter\let\csname equation*\endcsname\relax
  \expandafter\let\csname endequation*\endcsname\relax
\usepackage{amsmath}
\usepackage{graphicx}
\usepackage[perpage]{footmisc}

\begin{document}

\title[Rydberg excitation of trapped ions ]{Towards Rydberg quantum logic with trapped ions}

\author{P.~Bachor, T.~Feldker, J.~Walz, and F.~Schmidt-Kaler} 
\address{QUANTUM, Institut f\"ur Physik, Johannes Gutenberg-Universit\"at Staudingerweg 7, and Helmholtz-Institut Mainz, Johann-Joachim-Becherweg 36, D-55128 Mainz, Germany} 
\ead{bachor@uni-mainz.de}

%The authors' affiliations follow the list of authors. Each address is set by using \verb"\address{#1}" with the address as the single parameter in braces. If there is more than one address then the appropriate superscripted number, followed by a space, should come at the start of the address. E-mail addresses are added by inserting the command \verb"\ead{#1}" after the postal address(es) where \verb"#1" is the e-mail address. See section~\ref{startsample} for sample coding. For more than one e-mail address, please use the command \verb"\eads{\mailto{#1}, \mailto{#2}}" with \verb"\mailto" surrounding each e-mail address.  Please ensure that, at the very least, you state the e-mail address of the corresponding author.

\vspace{10pt}

\begin{abstract}
	We demonstrate the excitation of ions to the Rydberg state $22F$ by vacuum ultraviolet radiation at a wavelength of 123\,nm combined with the coherent manipulation of the optical qubit transition in 
	$^{40}\text{Ca}^+$. With a tightly focused beam at 729\,nm wavelength we coherently excite a single ion from a linear string into the metastable $3D_{5/2}$ state before a VUV pulse excites it to the Rydberg state. In combination with ion shuttling in the trap, we extend this approach to the addressed excitation of multiple ions. The coherent initialization as well as the addressed Rydberg excitation are key prerequisites for more complex applications of Rydberg ions in quantum simulation or quantum information processing.
\end{abstract}

\pacs{37.10.Ty, 32.80.Ee, 42.50.Ex}
%37.10.Ty: Trapped ions
%32.80.Ee: Rydberg states
%42.50.Ex: Optical implementations of quantum information processing and transfer

% For two-column output uncomment the next line and choose [10pt] rather than [12pt] in the \documentclass declaration
%\ioptwocol

%\maketitle

\tableofcontents

\section{Motivation for Rydberg excitations in trapped ion crystals}
%Rydberg anregungen einbauen in die toolbox der qiv mit Ionen, denn wichtig für quant. comp., gatter und modendesigns für quant. sim.  
% auf das PRL hinweisen und darstellen, dass hier der schwerpunkt auf der kohärenten und räumlich addressierten Qubit anregung liegt, 
% Testübergang nach 22F

Trapped cold ion crystals have multiple applications in fundamental and applied science. This includes sophisticated processes in quantum simulation and quantum information applications \cite{Leibfried2003Rev, Blatt2012, Kielpinski2002}. Core elements are two-qubit quantum logic operations, implemented by spin-dependent light forces on ions, which are conveyed along the entire Coulomb crystal \cite{Cirac1995}. Quantum algorithms such as the Shor factorization algorithm \cite{Monz2015} and elementary quantum simulations of spin-dynamics \cite{Senko2015, Jurcevic2014} have been implemented. Such progress demonstrates the versatility of two-qubit gate operations, historically starting from the Cirac-Zoller gate \cite{Cirac1995, FSK2003}, the Moelmer-Soerensen gate \cite{Sorensen1999, Benhelm2008} or the geometric phase gate \cite{Leibfried2003}. On the other hand, impressive results are reported in the research area of Rydberg quantum logic operations \cite{Gaetan2009, Urban2009, Isenhower2010, Maller2015} and Rydberg many-body-physics, where the giant dipolar interactions are exploited to simulate quantum spin-models \cite{Barredo2015, Schauss2015, Schempp2015}, or investigate the role of impurities in quantum systems.  

Here, we present  our experimental approach to join the advantages of ion crystals for quantum information processing, thus implementing an optical addressable single qubit in a linear register, with Rydberg excitations and the corresponding giant dipolar forces. The interplay between Coulomb and Rydberg interactions will enable fast multi-qubit gate operations \cite{Li2013} and might also allow for the observation of novel many-body phenomena \cite{Nath2015} or non-equilibrium dynamics of such systems \cite{Li2012}.  

We exemplify our case by an addressed Rydberg excitation on the $3D_{5/2}$ to $22F$ transition in cold trapped Calcium ions at 123.256119(5)\,nm wavelength. The article starts describing the experimental apparatus which includes a segmented ion trap, and lasers to initialize and manipulate the optical qubit on the $4S_{1/2} \leftrightarrow 3D_{5/2}$ transition in $^{40}$Ca$^+$. We describe an optical assembly which allows for single ion addressing \cite{Nagerl1999} on the optical qubit transition, in combination with optical access for the VUV-laser beam which is used to drive the Rydberg excitation. We demonstrate the selective state preparation of single ions in a linear crystal, and the subsequent excitation to the $22F$ Rydberg state.  Recently, we have reported the excitation of higher lying $F$ states with principal quantum numbers $n>50$ and presented a systematic study of the effects of large electric polarizabilities for such states \cite{Feldker2015}. The combination of coherent manipulation of qubits with the excitation to Rydberg levels, as demonstrated in this publication, will be the key for designing the Rydberg-interactions and quantum gate operations. We discuss future challenges and plans in the conclusion. 

\section{Description of the experimental setup}
%zunächst einen Überblick geben zur Organistaion dieses Kapitels, dann
In this section we describe the experimental setup for the Rydberg excitation of $^{40}$Ca$^+$ ions. First, we explain the design of the segmented linear Paul trap and the integration of the laser beam at $729\,\text{nm}$ wavelength. Furthermore, we describe the generation of narrow band and continuous wave (cw) vacuum ultraviolet radiation near 123\,nm wavelength and the beam delivery to the ion crystal. 

\subsection{Linear segmented ion trap for Calcium ions }\label{sec:trap}
% Fallenaufbaubild
For the confinement of the ions a linear Paul trap with segmented electrodes is used. The blades for the segmented electrodes are fabricated from a 125\,$\mu$m thick waver of alumina substrate cut with femtosecond laser pulses. Ceramic parts get a 300\,nm evaporated gold coating. They are glued to a titanium frame such that they form a X-shaped linear trap with 480\,$\mu$m distance from electrode surfaces to the symmetry center. A radio frequency (RF) voltage is applied to two diagonally opposing blades to generate the radial potential. The RF voltage comes by a signal generator \footnote{HP-8657B Signal Generator}, an analog attenuator \footnote{mini circuits ZX73-2500} allows for fast adjustment of the RF-amplitude, followed by a power amplifier \footnote{mini circuits ZHL-5W-1, $5\, \text{W}$}. A helical resonator is used for impedance matching of the amplifier output with the ion trap circuit. We have operated two different helical resonators with resonance frequencies $\Omega_{\text{RF}}/2 \pi = 6.5\,$MHz or $\Omega_{\text{RF}}/2 \pi = 20\,\text{MHz}$, respectively. An amplitude of up to $U_{\text{RF}}= 200\,\text{V}$ results in radial vibrational secular frequencies $\omega_{\text{rad}}/2 \pi = 700\,\text{kHz} - 2\, \text{MHz}$. 

\begin{figure}[htbp]
	\centering
		\includegraphics[width=0.50\textwidth]{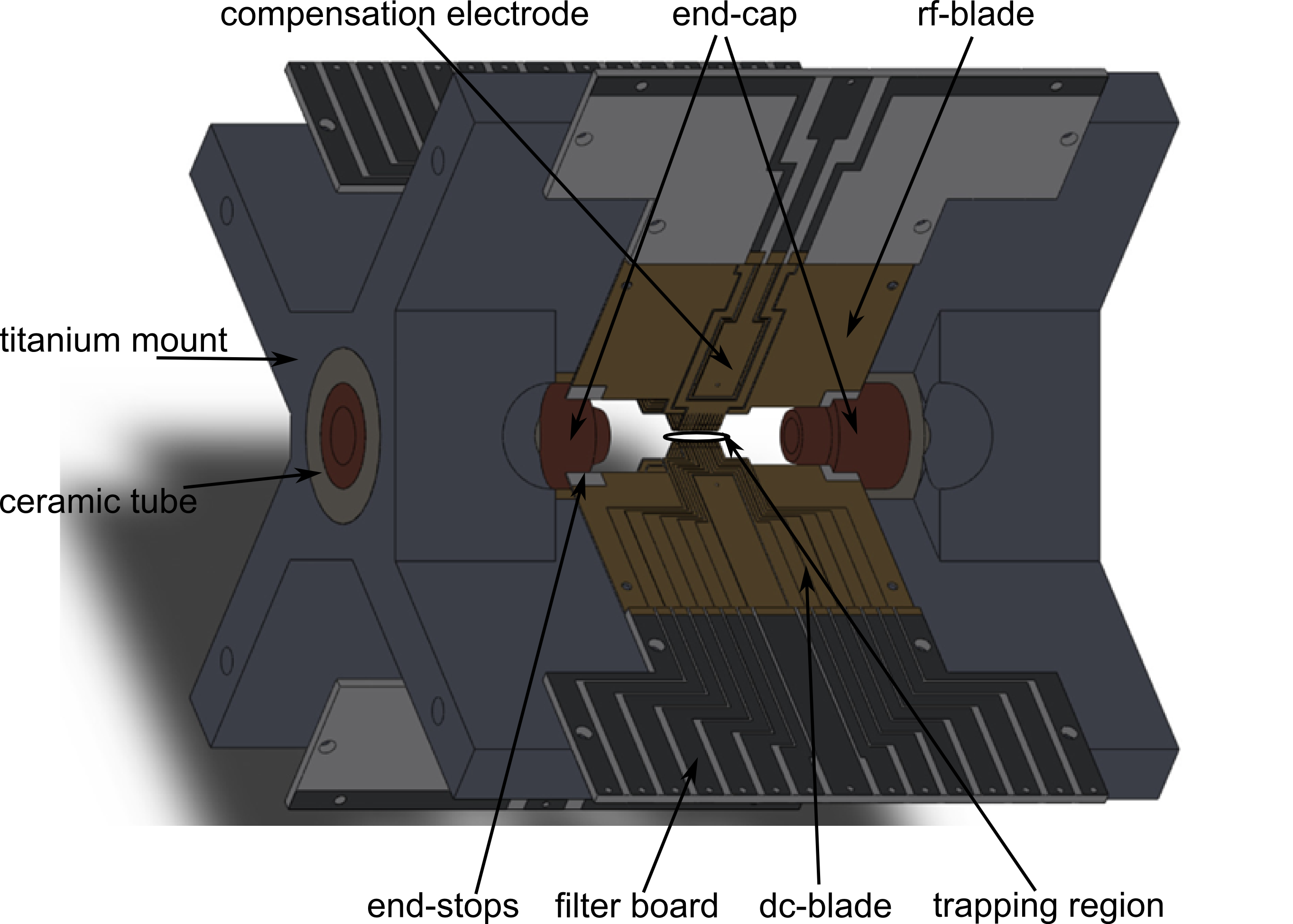}
	\caption{{\bf Construction drawing of the segmented Paul trap.} The trap electrodes are made from gold coated ceramic blades, glued to an X-shaped titanium mount. See text for more details.}
	\label{fig:Fig1_NewTrap3D}
\end{figure}

The trap is designed (Fig.~\ref{fig:Fig1_NewTrap3D} and Ref.~\cite{Jacob2014}) to optimize the accuracy of the radial alignment of all four blades with respect to the symmetry trap axis, thus minimizing micro-motion. Differential charging of the surfaces and residual imperfections of the device are compensated by correction voltages to center the ions at the RF-node. Compensation electrodes are placed behind the RF rails on the respective blades. We operate these electrodes with voltages of $\pm 10-100$\,V, generated by DC/DC converters\footnote{HighTek; PM12-100$\pm$} with additional low pass filters, controlled by the analogue output (AO) of a NI card \footnote{National instruments; PCI-6703}. 

The DC blades are segmented in 11 electrodes, each of width $180\, \mu \text{m}$, separated by a $30\, \mu \text{m}$ insulation gap. Over the 11 segments the total length of the DC-blades is 2.28\,mm. DC-voltages are fed to each of the electrodes, to generate the axial confinement. Additionally, two steel-cones, with a center-pierced hole of $750\,\mu$m radius are mounted at a distance of 12\,mm on the symmetry axis. They are electrically insulated and can act as DC-end caps of the trap. The holes provide axial optical access for the vacuum ultra violet (VUV) radiation. 

Each DC segment is connected via a CLC-filter with 50\,kHz cutoff frequency to a home-built multi-channel voltage source; a FPGA controlled device which delivers ultra low noise arbitrary waveforms in 4x12 parallel channels with an update rate of 400\,ns and a range of  $\pm 10 \text{V}$. This setup allows for the generation of customized axial potentials according to the various experimental requirements, as well as fast shuttling \cite{Walter2012} and separation \cite{Ruster2014} of ion crystals without thermal excitation. We generate axial potentials of $\omega_{\text{ax}}/2 \pi = 250\, \text{kHz} - 900\, \text{kHz}$.  

The quantization axis for the optical transitions is provided by a magnetic field with an angle of $\pi/4$ to the trap axis. The vacuum vessel admits for optical access parallel, perpendicular and with an angle of $\pi/4$ to the trap axis, see Fig.~\ref{trap}. We briefly describe the laser sources in the following. The vessel is evacuated by a NEXTorr~D~100-5 pump to about $10^{-11}$\,mbar. 

\subsection{Laser sources for optical cooling and detection}

Calcium atoms from a resistively heated oven are ionized with light near 423\,nm and 375\,nm. Ions are excited with light-fields near wavelengths of 393\,nm, 397\,nm, 854\,nm and 866\,nm. We employ grating stabilized diode lasers in Littrow configuration. Two commercially available Toptica DL pro lasers at wavelength of 393\,nm\footnote{LD-0395-0120-1, 13mW; Toptica} and 397\,nm\footnote{LD-0397-0030-1, 13mW; Toptica} and two home built lasers at 854\,nm\footnote{HL8342MG, 50mW, 852nm; opnext} and 866\,nm\footnote{LD-0870-0100-2, 100mW; Toptica} wavelength provide an optical output power of about 10\,mW each. All wavelengths are monitored by a wavelength meter\footnote{wavelength meter WSU-10 (infrared) and wavelength meter WSU-7 (blue); Toptica}. The lasers at 397\,nm and 866\,nm wavelength are stabilized to a medium finesse cavity using a Pound-Drever-Hall (PDH) locking scheme. A laser line-width well below 1\,MHz is reached. 

\begin{figure}[htbp]
\centering
\includegraphics[width=0.7\textwidth]{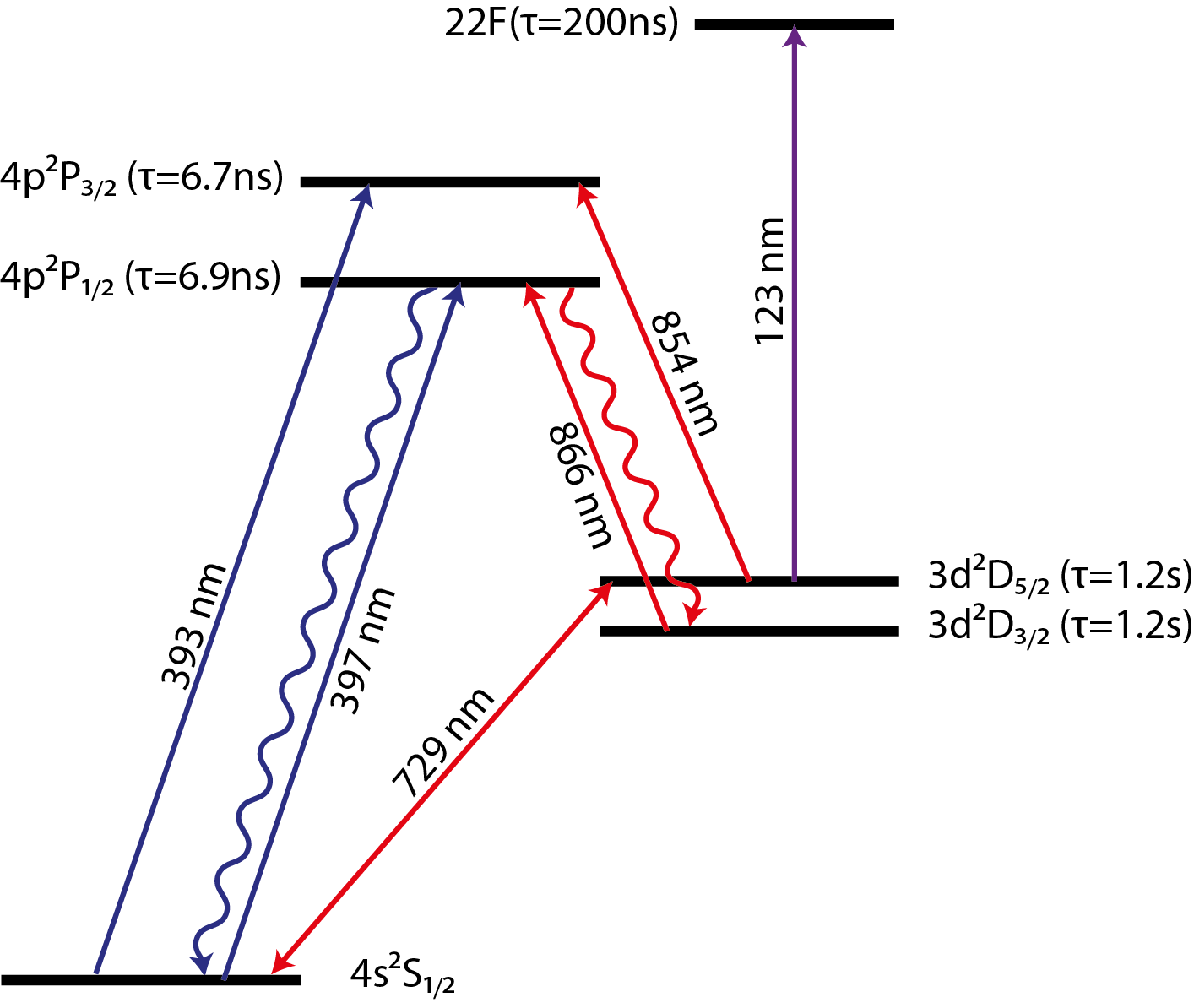}
\caption{\textbf{Levels and transitions in Calcium}
Laser sources near 432\,nm and 375\,nm are used for the two photon ionization of Calcium, sources near 393\,nm, 397\,nm, 866\,nm, and 854\,nm are employed for driving the dipole transitions in Ca$^+$, see text for details. The Rydberg state is excited from the long lived $3D_{5/2}$ state.}
\label{transitions}
\end{figure}

The laser beams are switched by acusto-optic modulators\footnote{BRI-QFZ-80-20-395 (blue) and BRI-TEF-80-20-860 (infrared); Brimrose} (AOMs) ensuring fast switching times $ < 1\,\mu$s and high repetition rates needed for the experiment. The light is delivered to the experiment by single mode polarization maintaining fibers and the beams are focused onto the ions with the fiber coupling optics. Relevant levels and transitions of $^{40}$Ca$^+$ are shown in Fig.~\ref{transitions}.

Ions are detected by imaging their fluorescence at 397\,nm wavelength to an electron multiplied charged coupled device (EMCCD) camera\footnote{Andor; IXon3 897; 512 $\times$ 512 pixel; pixel size $16\times16\,\mu$m}. A magnification of $m =17.7(1.4)$ of the imaging system is chosen and we fully resolve the fluorescance of individual ions in linear and planar ion crystals, where the inter-ion distances are a few $\mu$m.

\begin{figure}[htbp]
\centering
\includegraphics[width=0.7\textwidth]{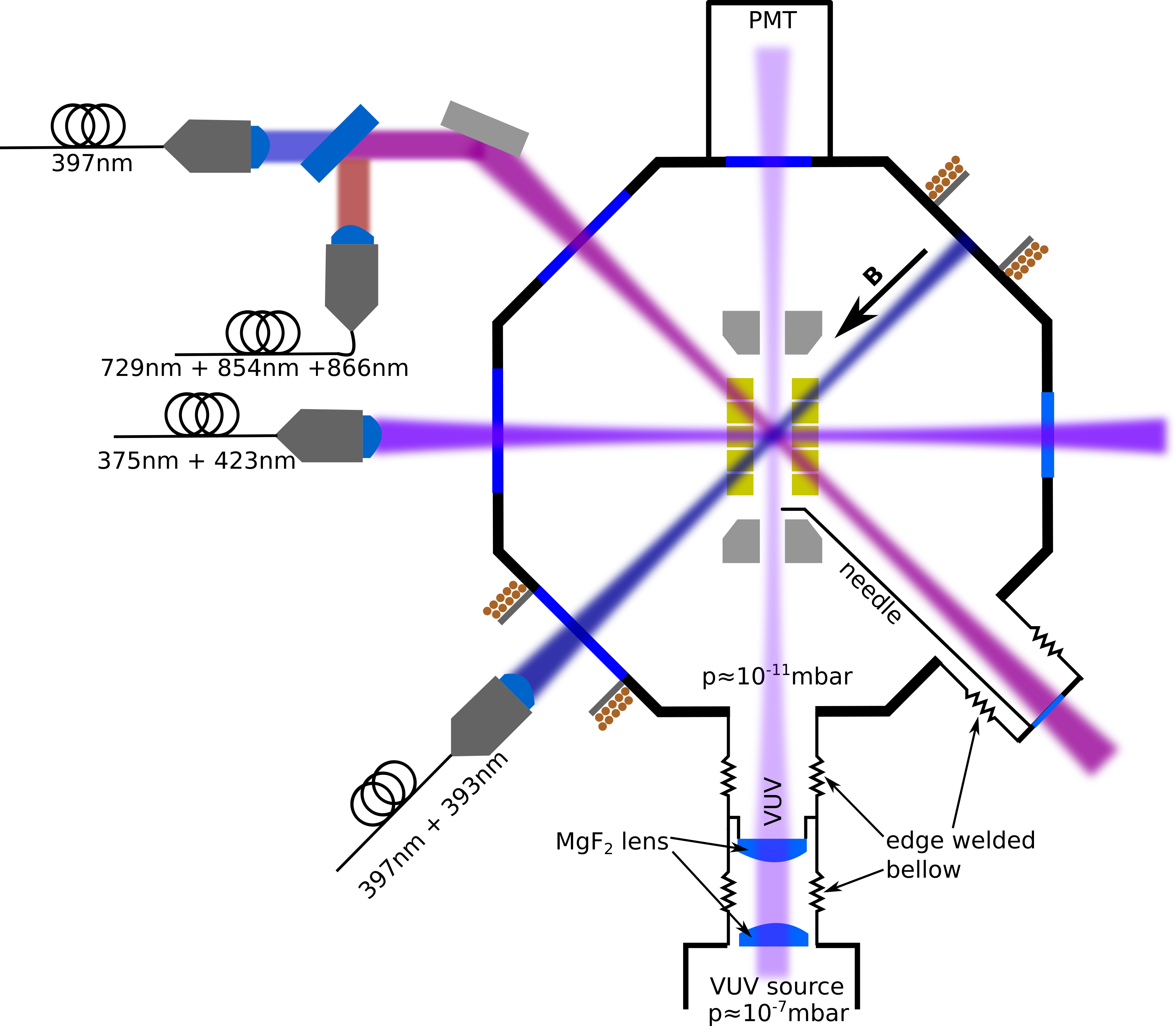}
\caption{\textbf{Optical access of the Paul trap.}
Directions of laser-beams and magnetic field in the ion trap viewed from the the top. The VUV beam propagates axially through the trap, focused by two MgF$_2$ lenses. Beams for Doppler cooling and optical pumping are delivered to the vacuum vessel by optical fibers and propagate parallel and perpendicular to the magnetic field, featuring a projection on the x, y, and z axis of the trap. The light at 729\,nm wavelength is superimposed with the Doppler beam, in the beam path (a) and used for globally illuminating the ion crystal with coupling to all motional modes.}
\label{trap}
\end{figure}

\subsection{Individual ion addressing on the optical qubit transition}

\begin{figure}[htbp]
\centering
\includegraphics[width=0.7\textwidth]{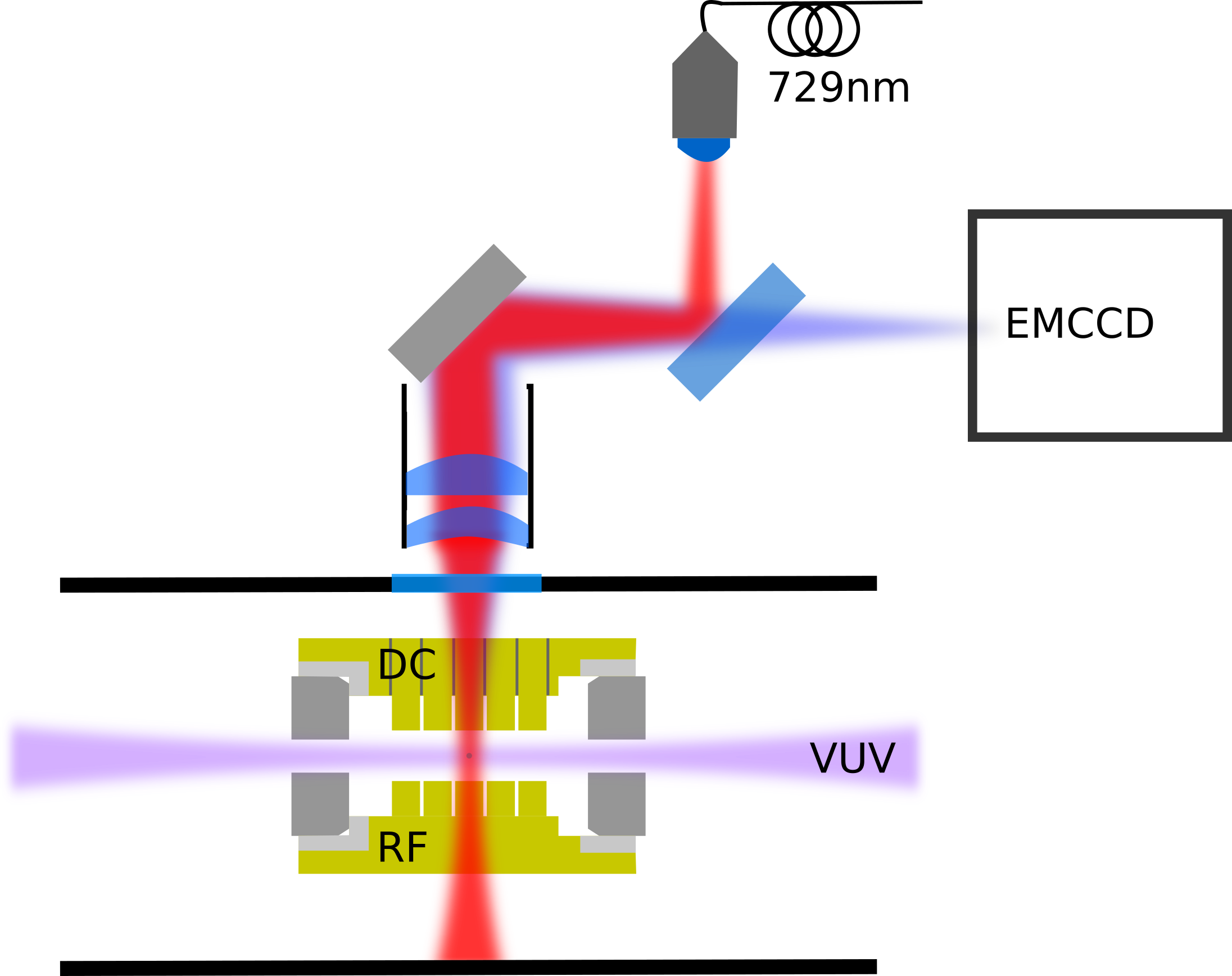}
\caption{\textbf{The light path of the single ion addressing beam.}
The light at 729nm wavelength is irradiated vertically into the trap on the beam path (b), superimposed on the fluorescence image, tightly focused by a lens with high numerical aperture. The fluorescence image is detected with an EMCCD camera. The two end caps allow for precise radial alignment of the blades. The segmented electrodes are separated by insulation gaps.}
\label{729top}
\end{figure}

The qubit information is carried in a single ion in two long-lived electronic states, namely the ground state and the metastable $3D_{5/2}$ level. To manipulate the optical qubit transition in $^{40}$Ca$^+$, we employ a Titanium:Sapphire laser \footnote{Matisse TX; Sirah , pumped by Millenia Pro 15s; Spectra Physics}, operated at 729\,nm wavelength and frequency stabilized \cite{Macha2012} to a high finesse ultra-stable reference cavity\footnote{6020 notched cavity $F=140000$; ATFilms}. We estimate a linewidth of better than 100\,Hz. The spectrally narrow radiation is used for resolved sideband cooling as well as individual coherent excitation of ions. The light is delivered to the experiment by a 100\,m long single mode fiber, hence fiber noise cancellation is employed to preserve the spectral purity of the light. We use an AOM operated at $110\pm 15\, \text{MHz}$, in double pass configuration for switching, pulse shaping and frequency tuning of the light. Two different beam paths (denoted (a) and (b)) deliver the light pulses to the ion crystal. For (a), the beam is coupled into the same fiber as the IR beams (854~nm and 866~nm), propagates through the trap perpendicular to the magnetic field and has a projection on all principal axes of the trapping potential. With a spot size of $50\,\mu m$ this beam illuminate the whole ion crystal. Rabi frequencies in the order of $\Omega/2 \pi = 100\, \text{kHz}$ are achieved with an optical power of 15\,mW. For path (b) the 729\,nm radiation is coupled into a separate single-mode and polarization-maintaining fiber, overlapped with the ion's fluorescence image, and focused\footnote{Sill Optics; focal length f=66.8mm, aperture D=38mm, for 397nm and 729nm wavelength} to a spot size of $\omega_0=4\, \mu\text{m}$. Due to the high intensity in the tight focus we achieve Rabi frequencies around $\Omega/2 \pi = 1 \text{MHz}$. This beam propagates vertically, perpendicular to the trap axis and is used to address single ions, see Fig.~\ref{729top}. We will describe the corresponding experimental results in Sect.~\ref{KapAdressing}.
 
\subsection{Vacuum ultra violet source near 123\,nm}
%sepz. das was noch nicht beschrieben war, gelber laser, linienbreiten
% 123nm Laser Bild 

The coherent VUV light near 123\,nm wavelength is generated in a nonlinear four-wave mixing process in mercury vapor. A triple-resonant scheme is implemented; laser beams at 254\,nm, 408\,nm and 580\,nm wavelength are superimposed via dichroic mirrors and focused into a mercury vapor cell. The mixing efficiency $\epsilon\propto|\chi^{(3)}|^2$ ranges between 20 and 30$\,\mu\text{W}/\text{W}^3$ at the required excitation wavelength, where $\chi^{(3)}$ denotes the third order nonlinear susceptibility. For a strong enhancement of $\chi^{(3)}$ the frequencies of the three fundamental radiation fields are tuned in the vicinity of atomic resonances in mercury \cite{Kolbe2012}. 

\begin{figure}[htbp]
\centering
\includegraphics[width=0.7\textwidth]{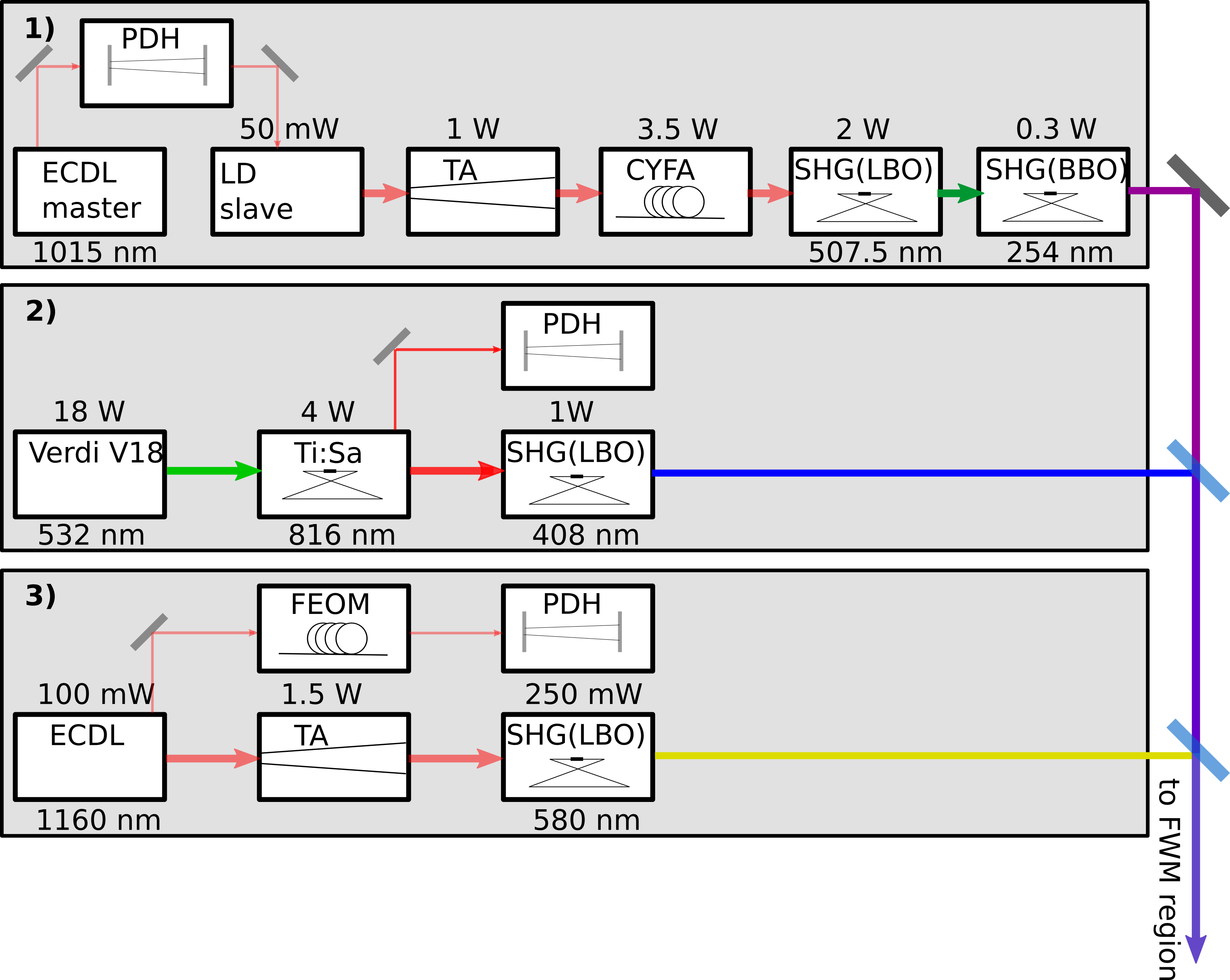}
\caption{\textbf{Block diagram of the lasers used to create the fundamental light fields at 254~nm (top), 408 nm (middle), and 580\,nm (bottom) wavelength for the FWM process.} The light is generated by frequency-doubling (quadrupling) the output of high power infrared lasers. Frequency stabilization to medium finesse cavities ensures narrow spectral bandwidth of the outcoming light. ECDL denotes the external cavity diode lasers, TA stands for tapered amplifier, the frequency servo uses a Pound-Drever-Hall (PDH) scheme, (C)YFA denotes the (cryogenic) ytterbium fiber amplifier and frequency doubling (SHG).}
\label{lasersetup}
\end{figure}

A schematic illustration of the laser setup is shown in Fig.~\ref{lasersetup}.
The ultraviolet light at 254\,nm wavelength is produced by the amplified and frequency-quadrupled output of a grating stabilized laser diode at 1015\,nm wavelength. We employ a master/slave diode laser system: The master diode's output is stabilized to a medium finesse cavity, the transmitted light is used for injection-locking of the slave diode. Hence, a narrow linewidth of about $ 60\,\text{kHz}$ is achieved. After the cryogenic fiber amplifier\,\cite{Steinborn2013} and the two frequency doubling stages we end up with a typical power of about 200\,mW. The beam is focused to a waist size of $\omega_0\approx 6\,\mu\text{m}$ in the center of the mercury mixing region.

The output of a stabilized Titanium:Sapphire laser at 816\,nm wavelength is frequency-doubled, to generate 1\,W of light at 408\,nm wavelength. We obtain a linewidth of about $30\,\text{kHz}$ and measure a spot size of $\omega_0\approx 14\, \mu \text{m}$ at the center of the mixing cell. 

These two fundamental light fields are fixed at an optimal wavelength for the nonlinear process. The third fundamental is tuned such that the resulting VUV wavelength matches the resonance conditions to excite the Rydberg transition in $^{40}$Ca$^+$. A high power laserdiode\footnote{Innolume GC-1160-90-TO-200-A} at 1160\,nm wavelength is operated in an external cavity in Littrow configuration. A high bandwidth fiber EOM\footnote{Photline NIR-MPX-LN-10 driven by R\&S SMP03 signal generator} modulates the light, generating sidebands at a computer controlled frequency. Stabilizing the frequency of one sideband to the resonance of the cavity allows for mode-hop-free frequency scanning of the laser in a range of $\pm$ 1\,GHz, and results in a spectral width of $\approx 80\, \text{kHz}$. About 80\,mW of this light is injected into a tapered amplifier\footnote{BoosTA pro, 1.5\,W}. For frequency doubling we use a LBO crystal in an enhancement bow-tie cavity. We generate an output power of 250\,mW at 580\,nm wavelength. The beam waist in the mercury vapor is determined as $\omega_0\approx 13 \, \mu \text{m}$ for optimized VUV generation. 

 We determine frequencies of all fundamental waves with a wavelength meter\footnote{HighFinesse WSU 10}. The frequency fluctuation of the generated VUV light is the sum of the fundamental spectral linewidths, resulting in a linewidth $\Gamma_{VUV}<500\,\text{kHz}$. We monitor the power of the VUV light transmitted through the trap with a photomultiplier. 

\subsection{Vacuum ultra violet beam delivery to trapped ion crystal}
%beschreibung des opt. aufbaus, fokusmessungen, justage methoden, stabilität und Grenzen der Fokusgröße  
A major challenge is the connection between the VUV source and the ion trap. Light at 123\,nm wavelength is absorbed by air, thus vacuum conditions are necessary for the complete beam path. The apparatus is sketched in Fig.~\ref{trap}. The beam is focused into the trap by a telescope which consists of two MgF$_2$ lenses. The second of the two lenses is glued airtight to a titanium mount in between two flexible bellows and separates the VUV generation zone with a pressure of about $10^{-7}$\,mbar from the ion trap's vacuum ($\sim 10^{-11}$\,mbar). 

For a coarse alignment of the VUV beam, the ion trap's vacuum vessel is moved with respect to the VUV source and the power transmitted through the trap is optimized. For fine-adjustment, we move a needle tip into the trap and observe the transmitted VUV power on a photomulitplier which is placed behind the trap. We determine the trap center by observing the ion's fluorescence on the EMCCD camera, and adjust the needle's position accordingly. The VUV beam is adjusted to the needle tip's position by moving the second lens of the telescope which is fixed to a 2D-translation stage. The VUV beam is focused into the trap  with $\omega_0\approx11\,\mu\text{m}$ passing the holes in the end caps.
Including transmission losses from optical elements in front of the trap we obtain $P_{VUV}\approx 10-80 \,\text{nW}$ at the ion.

\section{Experimental results on the $22F$ Rydberg excitation}

\begin{figure}[htbp]
\centering
\includegraphics[width=0.7\textwidth]{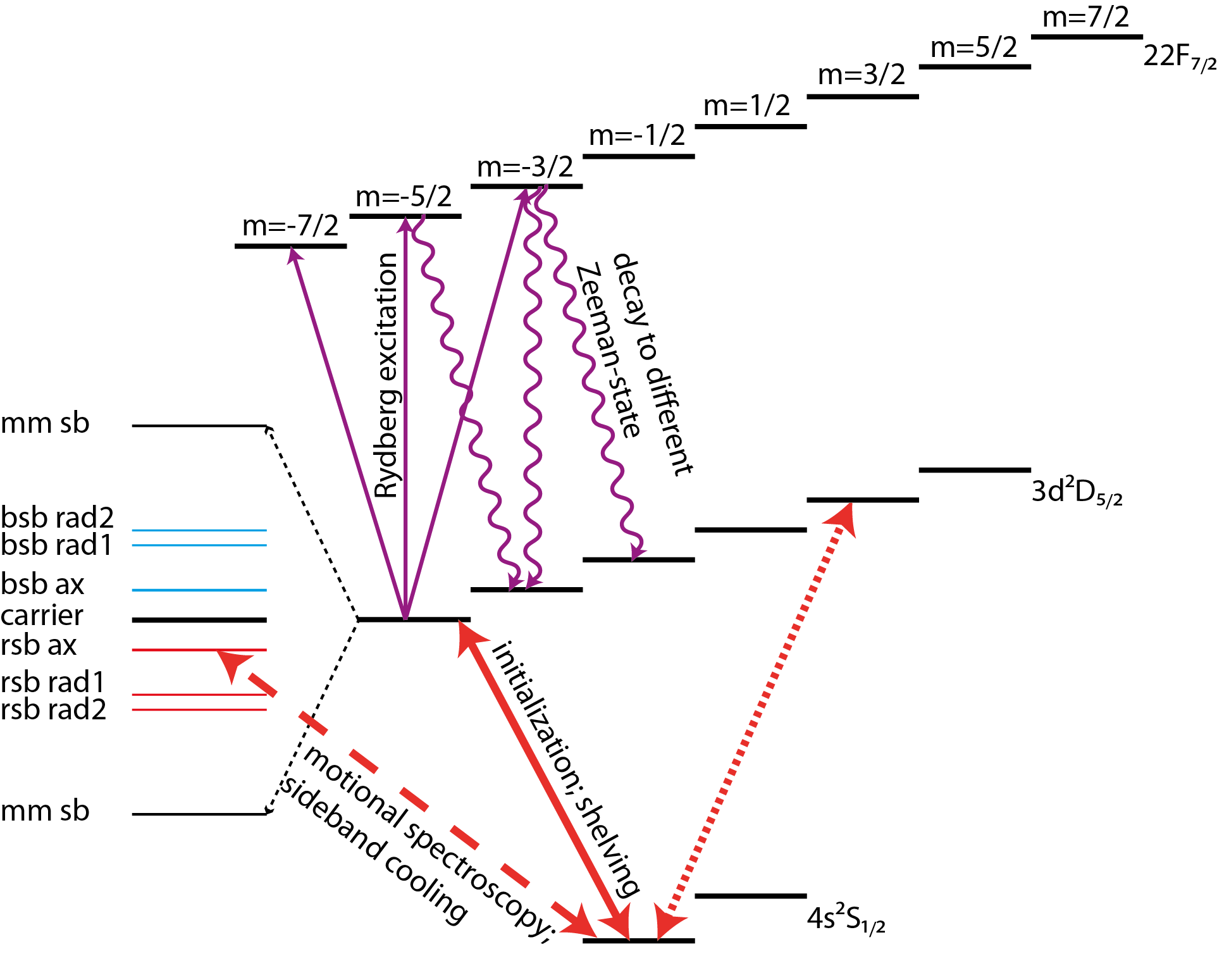}
\caption{\textbf{Zeeman structure of the energy levels used in the Rydberg excitation.} 
The ion is initialized by a $\pi$-rotation on the $\left|4S_{1/2}, -1/2\right\rangle \leftrightarrow \left|3D_{5/2},-5/2\right\rangle$ transition (solid red arrow). Alternatively the ion can be initialized in other Zeeman states (red dotted arrow). Individual Zeeman transitions are resolved due to the sharp resonance of this transition. Subsequently the ion is excited to the $22F$ state. Individual Zeeman transitions are not resolved in this step as the Rydberg resonance is broadened by the thermal motion of the ion and the linewidth of the VUV laser to about 3\,MHz. The excitation, followed by a decay back to the $3D_{5/2}$ state can change the Zeeman state. A second $\pi$-rotation is applied on the original $\left|4S_{1/2}, -1/2\right\rangle \leftrightarrow \left|3D_{5/2},-5/2\right\rangle$ transition, which rotates solely the population of the initial Zeeman state back to the ground state. Additionally the $4S_{1/2} \leftrightarrow 3D_{5/2}$ transition is used for precise spectroscopy of the magnetic field, as well as the thermal and driven (micro) motion of the ion (red dashed arrow). Resolved sideband cooling on this transition brings the ion close to the thermal ground state.}
\label{zeeman}
\end{figure}

In a recent publication \cite{Feldker2015} we have presented the excitation of Rydberg $F$ states with $n>50$ from the $3D_{3/2}$ state which has been populated by incoherent optical pumping. In the experiments presented here, however, we initialize the D-level by a coherent excitation on the $4S_{1/2} \leftrightarrow 3D_{5/2}$ transition. The upper qubit level is precisely the initial state for Rydberg excitation. In this section we present two examples for the toolbox opened up by the combination of Rydberg excitation of ions and coherent operation on long-lived electronically low lying qubit states: The first subsection attends to the initialization of the ion with a $\pi$-rotation into specific Zeeman states of the $3D_{5/2}$ followed by Rydberg excitation with VUV radiation. In the second part we demonstrate individually addressed Rydberg excitation of one and two ions in a linear Coulomb crystal. Beyond the initialization in the upper qubit level $3D_{5/2}$, the integration of narrow bandwidth radiation at 729\,nm into the setup has further benefits. Resolved sideband spectroscopy allows for a precise determination of the ions motional state, in regard to the thermal motion and the driven micro-motion (see Fig.~\ref{zeeman}). This is essential, as both effects do affect the line shape of the Rydberg transition. Furthermore, by using the narrow quadrupole transition near 729~nm for by resolved sideband cooling we can be prepare the ion motion close to the vibrational ground state.

Here, we combine Rydberg excitation to the $22F$ state at a wavelength of $\lambda_{22F}=123.256119(5)\,\text{nm}$ with coherent manipulation of the optical qubit in $^{40}$Ca$^+$. This is a key prerequisite for the use of Rydberg dipole-dipole interaction for generating multi-particle entanglement of ions. 

\subsection{Zeeman-state selective qubit initialization and Rydberg excitation} \label{sec:Zeeman}
% Spektr. und Rabi Flops auf Zeeman Zustände nur mit 729 carrier 

The magnetic field which we apply in the experiment lifts the degeneracy of Zeeman-states by $E_z = \mu_B \cdot g_j \cdot B $. Due to the narrow resonance of the $4S_{1/2} \leftrightarrow 3D_{5/2}$ transition, individual Zeeman-states can be excited. After side band cooling and optical pumping to the $\left|4S_{1/2},\pm1/2\right\rangle$ state, we reach about $90 \%$ population transfer to the desired $\left|3D_{5/2},m\right\rangle$ state, with a single $\pi$-rotation. In contrast to the sharp $4S_{1/2} \leftrightarrow 3D_{5/2}$ resonance, the resonance width for the transition to the Rydberg $22F$ state is much broader due to the comparatively short lifetime of this state ($\tau = 200\,$ns\,\cite{Glukhov2013}), the thermal motion of the ion and the linewidth of the VUV radiation. In the $F_{7/2}$ manifold a magnetic field of $B=0.28\,\text{mT}$ yields a frequency splitting of $4.5\,\text{MHz}$. Consequently we do not resolve the individual Zeeman-states of the $22F$, but the VUV radiation at $123.256119(5)\,\text{nm}$ wavelength induces transition with $\Delta m = -1,0,1$. Together with a subsequent decay back to the $3D_{5/2}$ state (again subjected to the selection rule $\Delta m=-1,0,1$), this results in a population transfer out of the initial Zeeman-state which indicates successful excitation. 

\begin{figure}[htbp]
\centering
\includegraphics[width=0.7\textwidth]{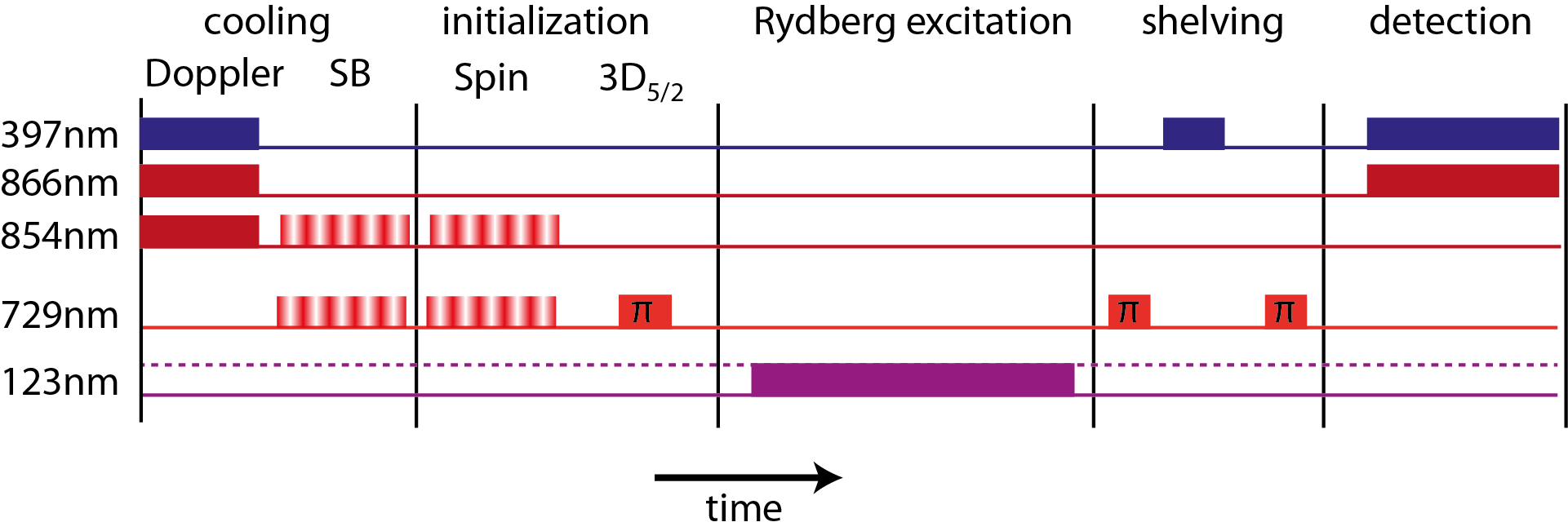}
\caption{\textbf{Pulse sequence used for excitation of different Zeeman states.}
During the initialization sequence the ions are excited into the $3D_{5/2}$ state via $\pi$-pulses. The shelving sequence prepares for the fluorescence detection of Rydberg excitation.}
\label{zeemanPulse}
\end{figure}

The new measurement sequence is based on fluorescence detection which allows for discriminating between the $3D_{5/2}$ state (no emitted photons - dark ion) and the $4S_{1/2}, 3D_{3/2}$ states (continuous fluorescence - bright ion) with high-fidelity. We use additional laser pulses to measure the population transfer between Zeeman levels, induced by the excitation to the Rydberg state. 
a) Ions are initialized with a $\pi$-pulse on the $\left|4S_{1/2},\pm1/2\right\rangle \leftrightarrow \left|3D_{5/2},\pm5/2\right\rangle$ transition. b) A VUV pulse of 1.5\,ms excites ions to the Rydberg states, from where they decay rapidly back to the $3D_{5/2}$. c) A $\pi$-pulse on the $\left|4S_{1/2},\pm1/2\right\rangle \leftrightarrow \left|3D_{5/2},\pm5/2\right\rangle$ transition rotates population that did not change the Zeeman state back to the ground state. d) Optical pumping with light at 397\,nm, hides the ground state population in the $3D_{3/2}$ state. e) An additional pulse on the $\left|4S_{1/2},\pm1/2\right\rangle \leftrightarrow \left|3D_{5/2},\pm5/2\right\rangle$ transition increases the fidelity of the shelving further to $>95\%$. f) Light at wavelength of 397\,nm and 866\,nm is switched on for the fluorescence detection. The complete sequence is shown in Fig~\ref{zeemanPulse}. As compared to the detection technique which we used in our previous experiments\,\cite{Feldker2015}, were we observe a change in the total angular momentum $j$ as the Rydberg state is excited and decays back to the D state, this new method is more efficient by about a factor of 5. This enabled us to reduce the VUV-pulse time to 1.5\,ms.
  
\begin{figure}[htbp]
\centering
\includegraphics[width=0.7\textwidth]{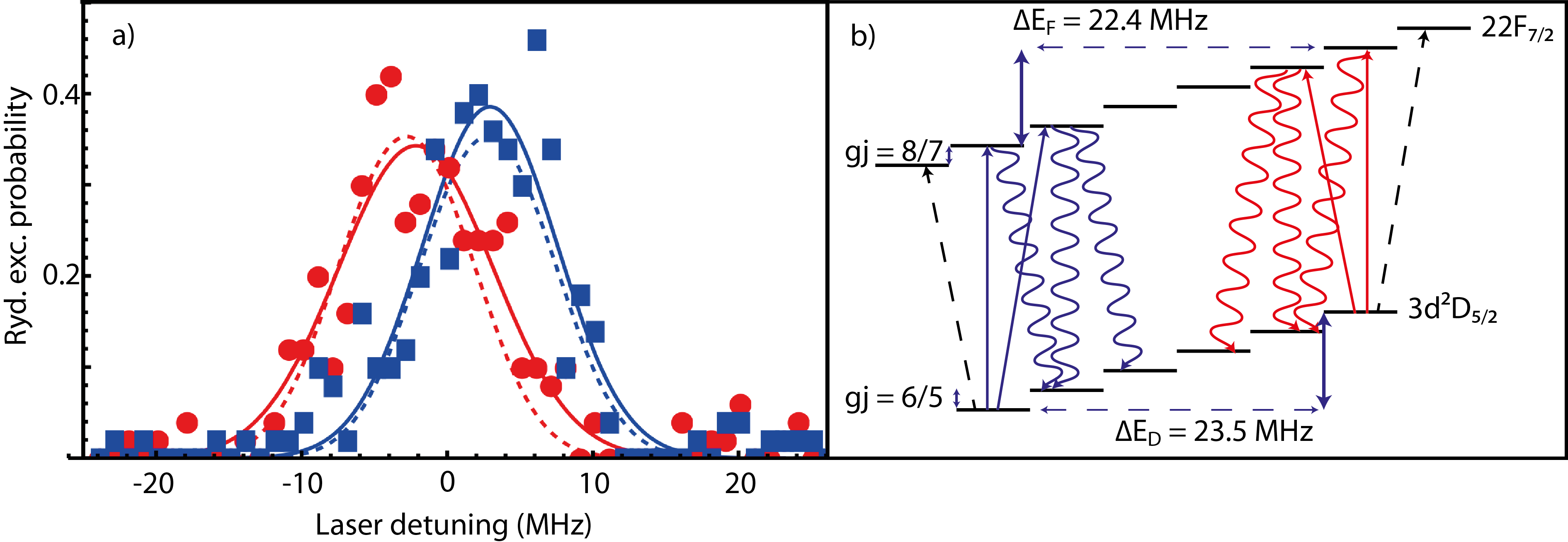}
\caption{\textbf{Experimental data of the excitation resonance of the $22F$ state from $3D_{5/2}$ $m = - 5/2$ (blue) and $m = 5/2$ (red).} a) Solid lines represent a fit of a Gaussian distribution to the experimental data, the linewidth from the fit are $\sigma = 5.3\,$MHz (red) and $\sigma = 4.6$\,MHz (blue) respectively. The separation of 5\,MHz agrees well with a model function (dashed lines) b). Illustration of the possible decay channels, after excitation from  $\left|3D_{5/2}, -5/2\right\rangle$ (blue) and $\left|3D_{5/2}, 5/2\right\rangle$ (red). The excitation to $\left|22F,\pm7/2\right\rangle$ can not be detected as the only decay channel is back to the initial state.}
\label{zeemanMeasurement}
\end{figure}

The data presented in Fig.~\ref{zeemanMeasurement}a shows the transition resonances for either the initial state $\left|3D_{5/2},+5/2\right\rangle$ or $\left|3D_{5/2},-5/2\right\rangle$. The resonances are composed of transitions to two Zeeman states each, with $\Delta m = \{0,1\}$ and $\Delta m = \{-1,0)\}$(see Fig~\ref{zeemanMeasurement}b). The observed frequency shift of 5\,MHz corresponds to the asymmetric excitation and detection scheme and an additional contribution of 1.1\,MHz from the different $g_j$ factors of $g_j(D_{5/2})= 6/5$ and $g_j(F_{7/2})=8/7$. The experimentally determined resonances agree with a model function consisting of Gaussian distributions with $\sigma = 3.8$\,MHz for each Zeeman transitions if we take into account a magnetic field of $B=0.28\,$mT. The strength of the magnetic field has been precisely measured by spectroscopy on the $4S_{1/2} \leftrightarrow 3D_{5/2}$ transition.

%zunächst einen Überblick geben zur Organistaion dieses Kapitels, dann
\subsection{Addressing single ions in a linear crystal}
\label{KapAdressing}

For advanced applications in quantum simulation or quantum computing it will be necessary to excite specific ions out of large Coulomb crystals. Here, we demonstrate the single ion addressing of Rydberg excitation. Instead of addressing single ions with a focused and steered VUV-beam, we initialize selected ions to the $3D_{5/2}$ state with the tightly focused beam at 729\,nm wavelength - only those ions will undergo excitation to the Rydberg level, even though the beam near 123\,nm is illuminating the entire ion crystal. This concept has important advantages: The $3D_{5/2}$ state lives for $\tau=1168(9)\,\text{ms}$\,\cite{Kreuter2005} such that we have sufficient time for pulses and transport or separation sequences to initialize and configure multiple ions before a single VUV pulse excites all D-state ions at once. Furthermore, it appears to be technically straight forward to address ions with a laser beam at 729\,nm wavelength while it would be much more challenging to rapidly steer on single ions with a VUV beam. 

For this experiment we changed the trap drive radio frequency to $\Omega_{RF}/2\pi = 20\,$MHz (see Sect.\,\ref{sec:trap}) in order to generate a more stable potential for multiple ions. Vibrational frequencies of $\omega_{rad}/2\pi = 1.5$\,MHz and $\omega_{ax}/2\pi = 600$\,kHz where obtained, which corresponds to an inter ion distance of $d \approx 7\,\mu$m. 

\begin{figure}[htbp]
\centering
\includegraphics[width=0.7\textwidth]{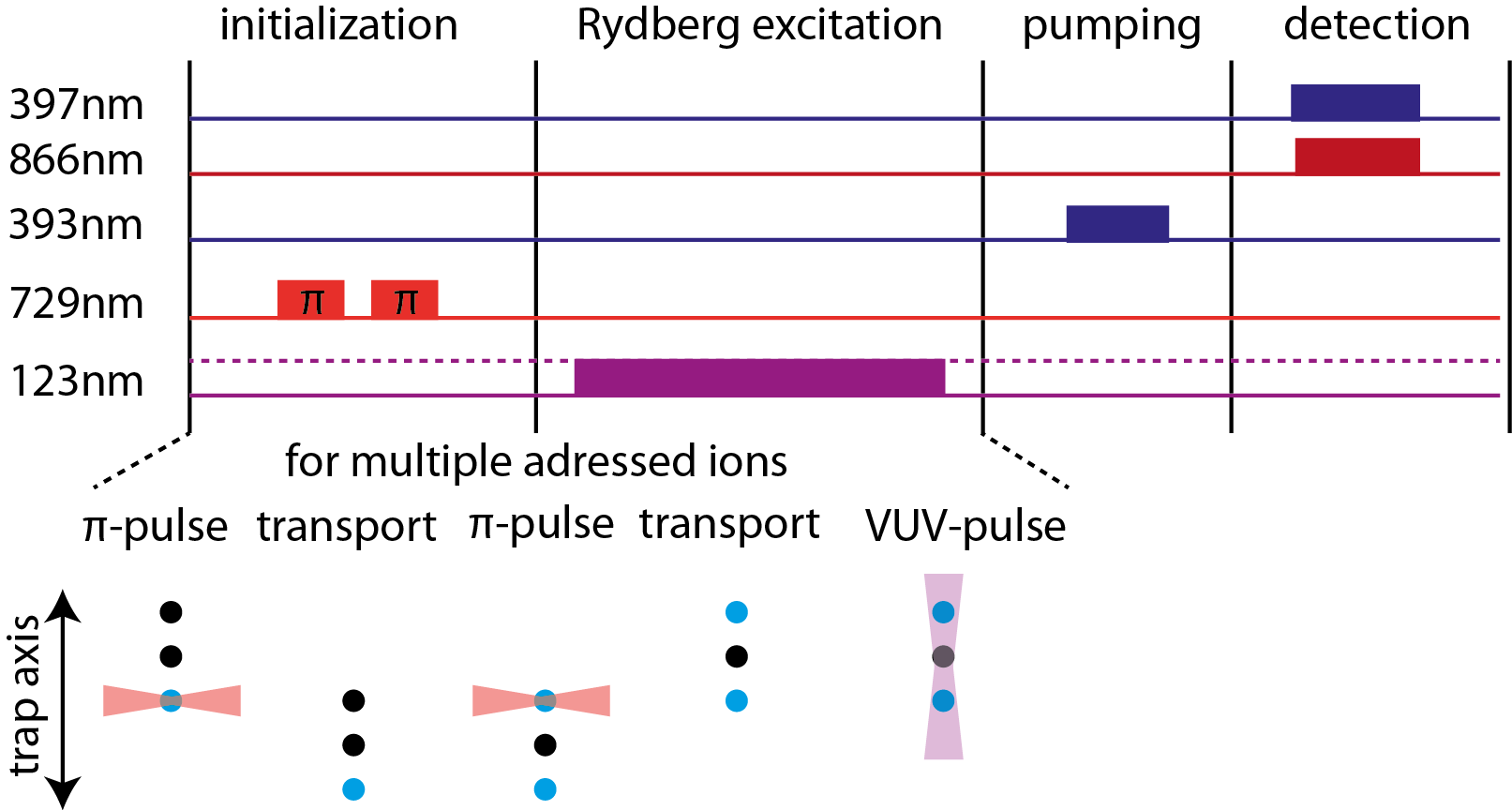}
\caption{\textbf{Pulse sequence for the Rydberg excitation and detection of selected ions.} In case of addressed excitation of multiple ions, the ions are initialized serially, applying a transport sequence between each pulse. The following Rydberg excitation is done on all initialized ions at once.}
\label{pulsSeqAddress}
\end{figure}

The pulse sequence used for the addressed excitation of one ion, or two ions, in a crystal of three ions is illustrated in Fig.\,\ref{pulsSeqAddress}. a) Two $\pi$-pulses on different Zeeman transitions are used for the initialization of each ion in order to increase the population transfer to $>90\%$. b) A VUV pulse of $\approx 5\,$ms excites the ions on the $3D_{5/2} \rightarrow 22F$ transition, a fraction of the excited ions decay to the $3D_{3/2}$ state. c) ground state population of all ions is pumped to the $3D_{5/2}$ state with 90\% fidelity. d) Thus, the initialized and the non-initialized ions are in the dark state ($3D_{5/2}$), only ions excited to Rydberg states and decayed to the $3D_{3/2}$ scatter photons during illumination with light at 397\,nm wavelength. 
With this scheme we avoid the use of addressed shelving pulses after the Rydberg excitation which could potentially distort the result. On the downside, only 90\% of the ground state population end up in the $3D_{5/2}$ state after pumping with light at 393\,nm wavelength. While this only has a small effect for the addressed ions as they are predominantly in the $3D_{5/2}$ or $3D_{3/2}$ state at this point, it leads to a background of 10\% for the non addressed ions to be in the 'bright' state. This effect is accounted for by adding 8\% constant background to the data of the addressed ions. 
%Note, that the fidelity of a $\pi$-pulse may be improved to better than 99$\%$, 

\begin{figure}[htbp]
\centering
\includegraphics[width=0.7\textwidth]{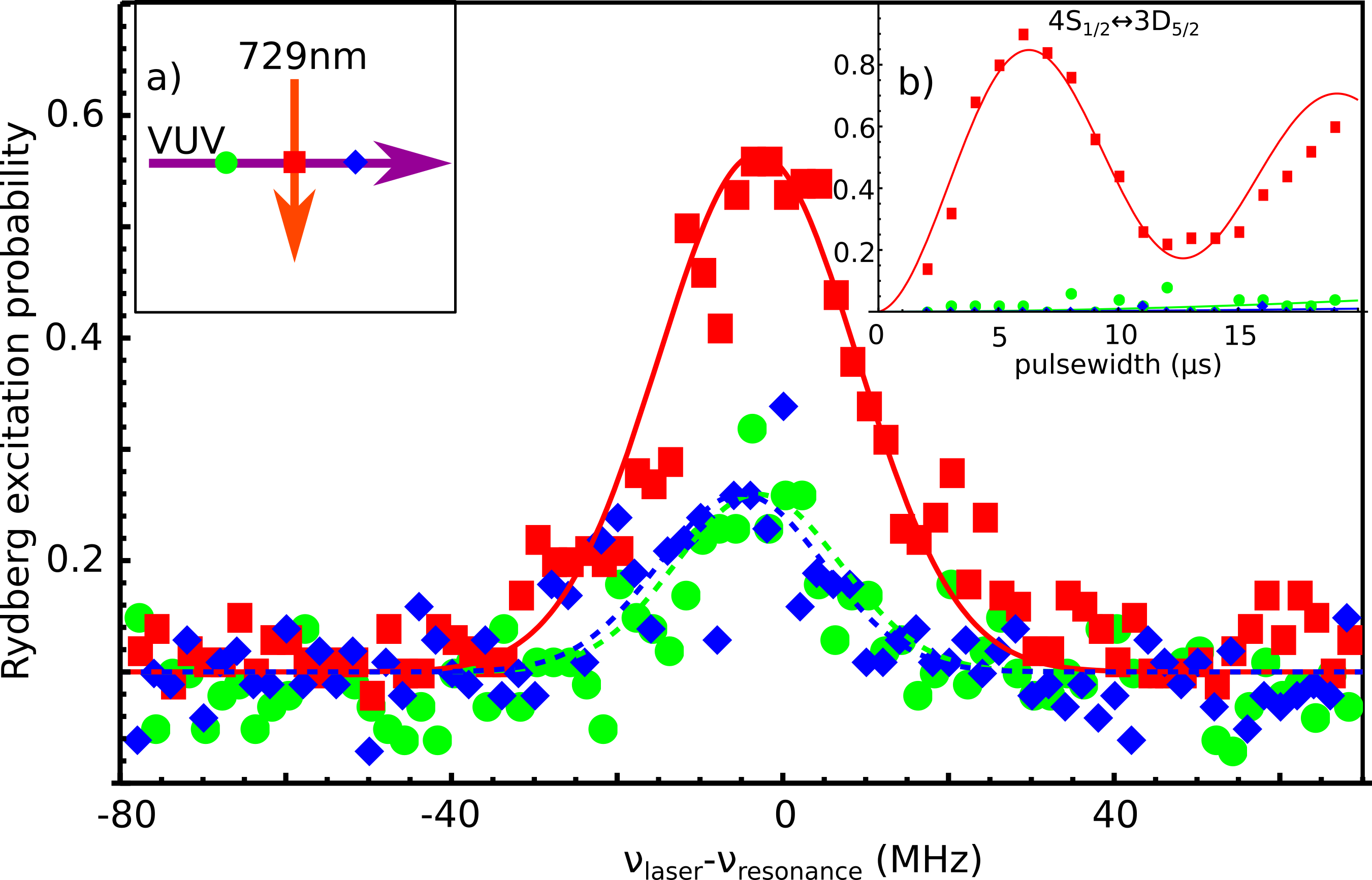}
\caption{\textbf{Addressed Rydberg excitation of the central ion out of a linear three ion crystal.} The ratio for Rydberg excitation on the addressed to the non addressed ions is $\approx 3$ with subtracted background signal. A constant background of 8\% was added to the data of the addressed ion to account for this systematic error (see explanation in text). Inset a) shows the geometry of the beams with respect to the ion chain. Inset b) shows addressed Rabi oscillations on the $4S_{1/2} \leftrightarrow 3D_{5/2}$ transition. The population of the central ion flops at a frequency of $\Omega_{\text{rabi}}/2\pi = 80\,\text{kHz}$ while the excitation of the outer ions remains negligible.}
\label{centralIon}
\end{figure}

The measured resonance with central ion addressed is shown in Fig.\,\ref{centralIon} the excitation rate on the addressed ion is about three times higher compared to the non addressed ion.The Rabi oscillations on the $4S_{1/2} \leftrightarrow 3D_{5/2}$ transition are shown in inset b) of Fig.\,\ref{centralIon}. During the time needed for a $\pi$-rotation on the addressed central ion, the excitation of the outer ions remains negligible. Still, residual Rydberg excitation of the non-addressed ions is observed. This behavior can be explained by the fact that we do not switch the VUV beam, but instead the VUV pulse width is determined by the time between initialization and detection. In contrast to the measurements presented in section\,\ref{sec:Zeeman}, all ions (addressed and non-addressed) which have not been excited to Rydberg states are in the $3D_{5/2}$ state during detection. Consequently they can be excited during detection and decay to a bright state. Switching the VUV beam with an electro optical modulator should solve this problem and allow for better addressing fidelity limited only by the quality of the initialization.

\begin{figure}[htbp]
\centering
\includegraphics[width=0.7\textwidth]{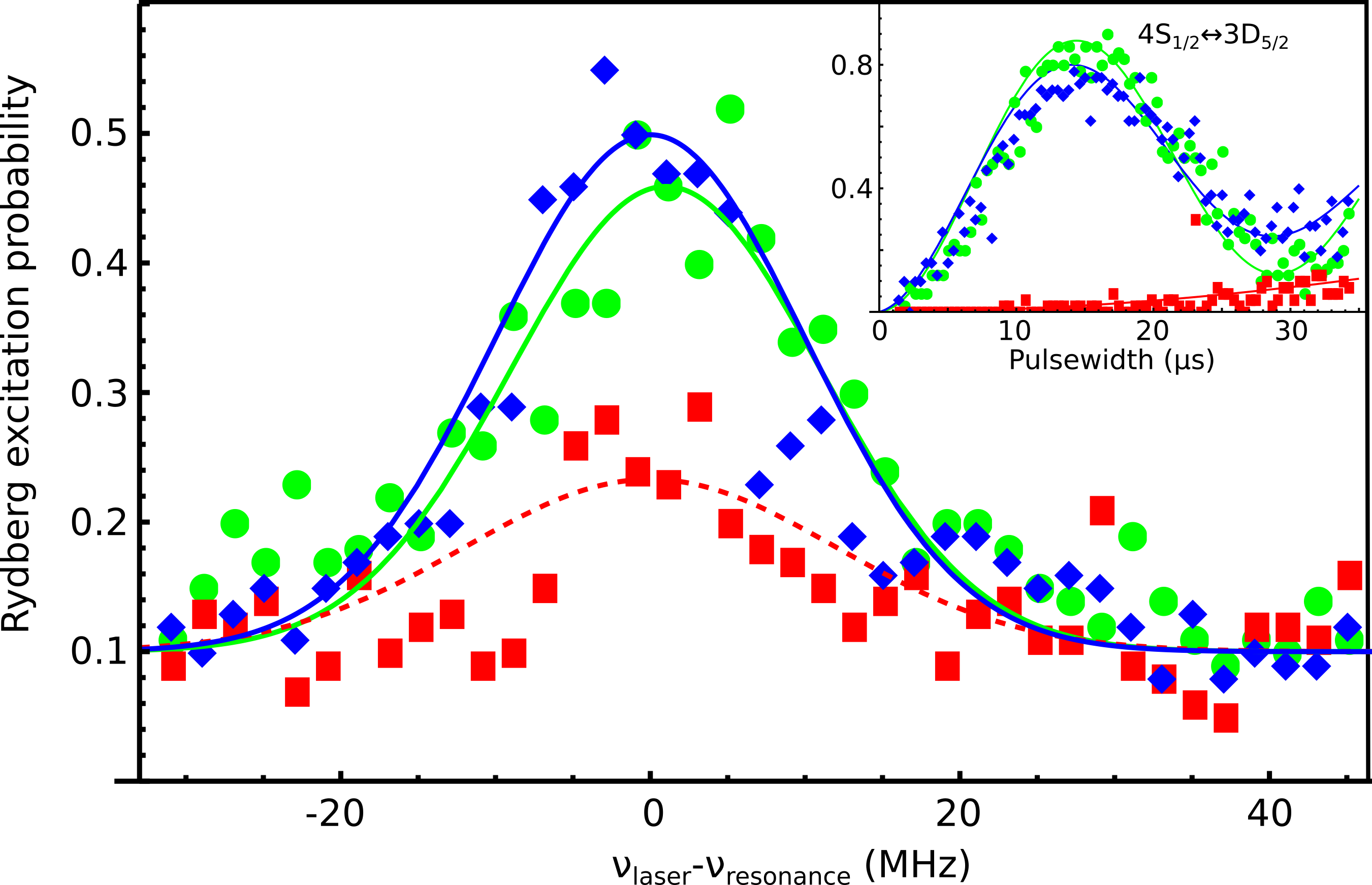}
\caption{\textbf{Addressed Rydberg excitation of both outer ions out of a linear three ion crystal.} The ratio for Rydberg excitation on the addressed to the non addressed ions is $\approx 2.8$ with subtracted background signal. Again, a constant background of 8\% was added to the data of the addressed ions. The inset shows the coherent population transfer on the addressed $4S_{1/2} \leftrightarrow 3D_{5/2}$ transition. Even with pulses on both outer ions the excitation of the non-addressed central ions remains negligible. Compared to the data for a single addressed ion, no significant loss in fidelity is introduced by the transport operation.}
\label{outerIon}
\end{figure}

Extending the scheme to the Rydberg excitation of multiple ions  is straight forward. The measured resonance with outer ions addressed is presented in Fig.\,\ref{outerIon}. In between laser pulses, the crystal is moved along the trap axis to superimpose the selected ion with the fixed 729\,nm beam. Voltage ramps of $\Delta V = \pm 280 \text{mV}$ on segments 4 and 8 have been used to displace the crystal by $14\,\mu$m in $500\,\mu$s. In inset b) the Rabi oscillations on the $4S_{1/2} \leftrightarrow 3D_{5/2}$ transition are shown. Compared to the sequence for a single addressed ion, neither the contrast of the oscillation for the addressed ion is reduced, nor the unwanted excitation of the non-addressed ion is increased by the transport operations. In principle this approach is scalable to long ion chains and even to two dimensional ion crystals as long as Rydberg excitation is restricted to ions on the trap axis. 

\section{Conclusion}

In conclusion, we have extended the wavelength range of  our VUV source in order to generate VUV radiation near 123\,nm such that we are able to excite the Rydberg state $22F$. We combined the excitation of Rydberg ions in Paul traps with the coherent manipulation of the optical qubit in $^{40}$Ca$^+$ which is an important step towards quantum simulation and quantum information applications with Rydberg ions. As a side effect, we increased the detection efficiency of the Rydberg excitation by detecting changes in the magnetic quantum number $m$ instead of the total angular momentum $j$. Using a segmented Paul trap, and the tightly focused beam at 729\,nm as well as the VUV beam at 123\,nm wavelength, we demonstrated the selective excitation of ions in a linear Coulomb crystal to Rydberg states. 

However, the excitation of the $22F$ state is only an intermediate step as its lifetime, polarizability and dipole interaction  between ions at a distance of a few $\mu$m would by far not allow Rydberg gate operations. We plan to excite long lived $nP$ states near $n \approx 50$ in order to combine the toolbox of coherent qubit manipulations with the unique features of interacting Rydberg ions in Paul traps.

The authors thank M. Stappel, D. Kolbe and R. Gerritsma for helpful discussions. The work was funded by the European Union H2020 FET Proactive project RySQ (Grant No. 640378). 

% Anregungslinie nP suchen, 
% schmale Linie und hohe Rabifrequenz
% auf Iogs PRA verweisen wegen MW-Dressing 

\hspace{2cm}

\paragraph{References}


\begin{thebibliography}{10}

\bibitem{Leibfried2003Rev}
D. Leibfried, R. Blatt, C. Monroe, and D. Wineland, {\it Rev. Mod. Phys.} {\bf 75,} 281 (2003).

\bibitem{Blatt2012}
R. Blatt, and C.~F. Roos, {\it Nature Physics} {\bf 8,} 277 (2012).

\bibitem{Kielpinski2002}
D. Kielpinski, C. Monroe, and D.~J. Wineland, {\it Nature} {\bf 417,} 709 (2002).

\bibitem{Cirac1995}
J.~I. Cirac, and P. Zoller, {\it Phys. Rev. Lett.} {\bf 74,} 4091 (1995).

\bibitem{Monz2015}
T. Monz, D. Nigg, E.~A. Martinez, M.~F. Brandl, P. Schindler, R. Rines, S.~X. Wang, I.~L. Chuang, and R. Blatt, arXiv:1507.08852. 

\bibitem{Senko2015}
C. Senko, P. Richerme, J. Smith, A. Lee, I. Cohen, A. Retzker, and C. Monroe, {\it Phys. Rev. X} {\bf 5,} 021026 (2015).

\bibitem{Jurcevic2014}
P.Jurcevic, B.~P. Lanyon, P. Hauke, C. Hempel, P. Zoller, R. Blatt, and C.~F. Roos, {\it Nature} {\bf 511,} 202 (2014).

\bibitem{FSK2003}
F. Schmidt-Kaler, H. H\"affner, M. Riebe, S. Gulde, G.~P.~T. Lancaster, T. Deuschle, C. Becher, C.~F. Roos, J. Eschner, and R. Blatt, {\it Nature} {\bf 422,} 408 (2003).

\bibitem{Sorensen1999}
A. S{\o}rensen, and K. M{\o}lmer, {\it Phys. Rev. Lett.} {\bf 82,} 1971 (1999).

\bibitem{Benhelm2008}
J. Benhelm, G. Kirchmair, C.~F. Roos, and R. Blatt, {\it Nature Physics} {\bf 4,} 463 (2008).
 
%\bibitem{Kirchmair2009}
%G. Kirchmair, J. Benhelm, F. Z\"ahringer, R. Gerritsma, C.~F. Roos, and R. Blatt, {\it New %Journal of Physics} {\bf 11,} 023002 (2009).

\bibitem{Leibfried2003}
D. Leibfried, B. DeMarco, V. Meyer, D. Lucas, M. Barret, J. Britton, W.~M. Itano, B. Jelenkovic, C. Langer, T. Rosenband, and D.~J. Wineland, {\it Nature} {\bf 422,} 412 (2003).

\bibitem{Gaetan2009}
A. Ga\"etan, Y. Miroshnychenko, T. Wilk, A. Chotia, M. Viteau, D. Comparat, P. Pillet, A. Browaeys, and P. Grangier, {\it Nature Physics} {\bf 5,} 115 (2009).

\bibitem{Urban2009}
E. Urban, T.~A. Johnson, T. Henage, L. Isenhower, D.~D. Yavuz, T.~G. Walker, and M. Saffman, {\it Nature Physics} {\bf 5,} 110 (2009).

\bibitem{Isenhower2010}
L. Isenhower, E. Urban, X.~L. Zhang, A.~T. Gill, T. Henage, T.~A. Johnson, T.~G. Walker, and M. Saffman, {\it Phys. Rev. Lett.} {\bf 104,} 010503 (2010). 

\bibitem{Maller2015}
K.~M. Maller, M.~T. Lichtman, T. Sia, Y. Sun, M.~J. Piotrowicz, A.~W. Carr, L. Isenhower, and M. Saffman, {\it Phys. Rev. A} {\bf 92,} 022336 (2015).

\bibitem{Barredo2015}
D. Barredo, H. Labuhn, S. Ravets, T. Lahaye, A. Browaeys, and C.~S. Adams, {\it Phys. Rev. Lett.} {\bf 114,} 113002 (2015).

\bibitem{Schauss2015}
P. Schau\ss, J. Zeiher, T. Fukuhara, S. Hild, M. Cheneau, T. Macri, T. Pohl, I. Bloch, and C. Gross, {\it Science} {\bf 347,} 1455 (2015).

\bibitem{Schempp2015}
H. Schempp, G. G\"unter, S. W\"uster, M. Weidem\"uller, and S. Whitlock, {\it Phys. Rev. Lett.} {\bf 115,} 093002 (2015).

\bibitem{Li2013}
W. Li, A.~W. Glaetzle, R. Nath, and I. Lesanovsky, {\it Phys. Rev. A} {\bf 87,} 052304 (2013).

\bibitem{Li2012}
W. Li, and I. Lesanovsky, {\it Phys. Rev. Lett.} {\bf 108,} 023003 (2012).

\bibitem{Nath2015}
R. Nath, M. Dalmonte, A.~W. Glaetzle, P. Zoller, F. Schmidt-Kaler, and R. Gerritsma, {\it New Journal of Physics} {\bf 17} (2015).

\bibitem{Nagerl1999}
H.~C. N\"agerl, D. Leibfried, H. Rohde, G. Thalhammer, J. Eschner, F. Schmidt-Kaler, and R. Blatt, {\it Phys. Rev. A} {\bf 60,} 145 (1999).

\bibitem{Feldker2015}
T. Feldker, P. Bachor, M. Stappel, D. Kolbe, R. Gerritsma, J. Walu, and F. Schmidt-Kaler, {\it Phys. Rev. Lett.} {\bf 115,} 173001 (2015).

\bibitem{Jacob2014}
G. Jacob, K. Groot-Berning, S. Wolf, S. Ulm, L. Couturier, U. G. Poschinger, F. Schmidt-Kaler, K. Singer,
arxiv.org:1405.6480, (2014).

\bibitem{Walter2012}
A. Walther, F. Ziesel, T. Ruster, S.~T. Dawkins, K. Ott, M. Hettrich, K. Singer, F. Schmidt-Kaler, and U. Poschinger, {\it Phys. Rev. Lett.} {\bf 109,} 080501 (2012).

\bibitem{Ruster2014}
T. Ruster, C. Warschburger, H. Kaufmann, C.~T. Schmiegelow, A. Walther, M. Hettrich, A. Pfister, V. Kaushal, F. Schmidt-Kaler, and U.~G. Poschinger, {\it Phys. Rev. A} {\bf 90,} 033410 (2014).

\bibitem{Macha2012}
T. Macha {\it Frequenzstabilisierung eines Titan-Saphir-Laser und Verbesserung von Qubits mit Ca+ - Ionen}, (Diploma thesis) Gutenberg Univ. Mainz (2012).

\bibitem{Kolbe2012}
D. Kolbe, M. Scheid, and J. Walz, {\it Phys. Rev. Lett.} {\bf 109,} 063901 (2012).

\bibitem{Steinborn2013}
R. Steinborn, A. Koglbauer, P.Bachor, T. Diehl, D. Kolbe, M. Stappel, and J. Walz, {\it Optics Express} {\bf 21,} 22693 (2013).

\bibitem{Glukhov2013}
I.~L. Glukhov, E.~A. Nikitina, and V.~D. Ovsiannikov, {\it Spectroscopy of Atoms and Molecules} {\bf 115,} 9 (2013).

\bibitem{Kreuter2005}
A. Kreuter, C. Becher, G.~P.~T. Lancaster, A.~B. Mundt, C. Russo, H. H\"affner, C. Roos, W. H\"ansel, F. Schmidt-Kaler, R. Blatt, and M.~S. Safronova, {\it Phys. Rev. A} {\bf 71,} 032504 (2005).

\end{thebibliography}
\end{document}